\documentclass[prl,aps,twocolumn,showpacs]{revtex4}
\usepackage{dcolumn}
\usepackage{graphicx}
\usepackage{amssymb}
\usepackage{amsmath}
\begin{document}

\title{Topological Aspects of Charge-Carrier Transmission across \\
Grain Boundaries in Graphene}

\author{Fernando Gargiulo}
\affiliation{Institute of Theoretical Physics, Ecole Polytechnique F\'ed\'erale de Lausanne (EPFL), CH-1015 Lausanne, Switzerland}
\author{Oleg V. Yazyev}
\affiliation{Institute of Theoretical Physics, Ecole Polytechnique F\'ed\'erale de Lausanne (EPFL), CH-1015 Lausanne, Switzerland}
\date{\today}

\begin{abstract}
We systematically investigate the transmission of charge carriers 
across the grain-boundary defects in polycrystalline graphene by 
means of the Landauer-B\"uttiker formalism within the 
tight-binding approximation. Calculations reveal a strong suppression
of transmission at low energies upon decreasing the density of 
dislocations with the smallest Burger's vector ${\mathbf b}=(1,0)$.
The observed transport anomaly is explained from the point of view of 
back-scattering due to localized states of topological origin. These
states are related to the gauge field associated with all
dislocations characterized by ${\mathbf b}=(n,m)$ with $n-m \neq 3q$
($q \in \mathbb{Z}$).
Our work identifies an important source of charge-carrier 
scattering caused by topological defects present
in large-area graphene samples produced by chemical 
vapor deposition.
\end{abstract}



\maketitle

Since its isolation in 2004 \cite{Novoselov04}, graphene has been
attracting ever-increasing attention due to its extraordinary physical 
properties and potential technological applications
\cite{Geim07,Katsnelson07,CastroNeto09}. 
Early research experiments on graphene have been performed using single
micrometer-scale samples obtained by micromechanical
cleavage. However, technological applications require manufacturing processes
that would allow for robust production at much larger scales, e.g. 
chemical vapor deposition technique \cite{Kim09,Li09,Bae10}.
Recent experimental studies have shown that such extended graphene
samples tend to be polycrystalline, i.e. composed of micrometer-size 
single-crystalline domains of varying lattice 
orientation \cite{Huang11,Kim11,An11}. Dislocations and grain 
boundaries are responsible for breaking the 
long-range order in polycrystals. 
These topological defects inevitably affect
all physical properties of graphene \cite{Yazyev10b,Grantab10,Gunlycke11,Ferreira11a,Wei12,VanTuan13,Radchenko13}. 
In particular, it has been
demonstrated experimentally that grain boundaries dramatically alter the
electronic transport properties of graphene \cite{Yu11,Tsen12,Koepke12}. 
Understanding the effect of topological defects on charge-carrier transport 
in graphene is crucial for technological applications of 
this material in electronics, clean energy and related domains.

According to the Read-Shockley model \cite{Read50}, grain boundaries
in two-dimensional crystals are equivalent to one-dimensional
arrays of dislocations. In graphene, the low-energy configurations 
of the cores of constituent dislocations are composed of pairs of 
pentagons and heptagons, with the resulting Burger's vector dependent 
on their mutual positions \cite{Yazyev10a,Liu10,Carpio08,Carlsson11}.
Remarkably, in certain
highly ordered grain-boundary structures the conservation of
momentum results in a complete suppression of the transmission of
low-energy charge carriers \cite{Yazyev10b}.
However, it is of paramount importance to understand the factors 
determining the charge-carrier transmission probability in a more 
general situation when no symmetry-related selection is present. 
For example, this is the case of grain boundaries with strongly perturbed 
periodic arrangement of dislocations, such as the ones observed in 
graphene grown by chemical vapor deposition \cite{Huang11,Kim11}. 

In this Letter, we report a systematic study of the charge-carrier 
transmission across grain boundaries in graphene by means of the 
Landauer-B\"uttiker approach. We find that 
the structural topological invariant of dislocations \cite{Nelson02}, 
the Burger's vector $\mathbf{b}$, plays a crucial role in determining 
transport properties. In particular, for the case of grain boundaries 
formed by the minimal Burger's vector $\mathbf{b}=(1,0)$ dislocations, 
we find an unexpected suppression 
of the transmission of low-energy charge carriers in the limit
of small misorientation angles (or, equivalently, small dislocation
densities). This counter-intuitive behavior is explained from the 
point of view of resonant back-scattering involving localized states of 
topological origin, which arise due to the gauge field created by dislocations
characterized by ${\mathbf b}=(n,m)$ with $n-m \neq 3q$ ($q \in \mathbb{Z}$). 
The $\mathbf{b}=(1,1)$ dislocations are shown to behave as ordinary scattering
centers and have very weak effects on the electronic transport. 


In our study we employ the nearest-neighbor tight-binding model
Hamiltonian
\begin{equation}
\label{TB}
	H = - t \sum_{\langle i,j \rangle} [ c_{i}^\dagger c_{j} + {\rm h.c.} ], 
\end{equation} 
where $c_{i}$ ($c_{i}^\dagger$) annihilates (creates) an electron at 
site $i$ and $\langle i,j \rangle$ stands for pairs of 
nearest-neighbor atoms. The hopping integral $t = 2.7$~eV is assumed 
to be constant \cite{endnote1}. 
Coherent transport across grain boundaries in
graphene is studied within the Landauer-B\"uttiker formalism which
relates the conductance $G(E)$ at a given energy $E$ to the  
transmission $T(E)$ as $G(E)=G_{0}T(E)$, with
$G_{0}=2e^{2}/h$ being the conductance quantum \cite{Buttiker85}.
The transmission is evaluated by means of the non-equilibrium 
Green's function approach using two-terminal device
configurations, with contacts represented by the semi-infinite ideal
graphene leads:
\begin{equation}
\label{Transmission}
T=\mathrm{Tr}[\Gamma_{\mathrm{L}}G_{\mathrm{S}}^{\dagger}\Gamma_{\mathrm{R}}G_{\mathrm{S}}].
\end{equation}
The scattering region Green's function $G_{\mathrm{S}}$
is calculated as
\begin{equation}
\label{GF}
G_{\mathrm{S}}=[E^{+}I-H_{\mathrm{S}}-\Sigma_{\mathrm{L}}-\Sigma_{\mathrm{R}}]^{-1}
\end{equation}
employing the coupling matrices $\Gamma_{\mathrm{L}\left(\mathrm{R}\right)}$ for the left (right) lead given by
\begin{equation}
\label{Couplingmatrices}
\Gamma_{\mathrm{L}\left(\mathrm{R}\right)}=i[\Sigma_{\mathrm{L}\left(\mathrm{R}\right)}-\Sigma_{\mathrm{L}\left(\mathrm{R}\right)}^{\dagger}].
\end{equation}
In these expressions $H_{\mathrm{S}}$ is the Hamiltonian for the scattering region, $\Sigma_{\mathrm{L(R)}}$ are the self-energies 
which couple scattering region to the leads and $E^{+}=E+i\eta I\;(\eta\rightarrow0^{+})$. The dependence of $T$, $G_{\mathrm{S}}$, $\Gamma_{\mathrm{L}\left(\mathrm{R}\right)}$ and $\Sigma_{\mathrm{L(R)}}$ on energy $E$ and transverse momentum $k_{||}$ is omitted for the sake of compact notation.


\begin{figure}
\includegraphics[width=86mm]{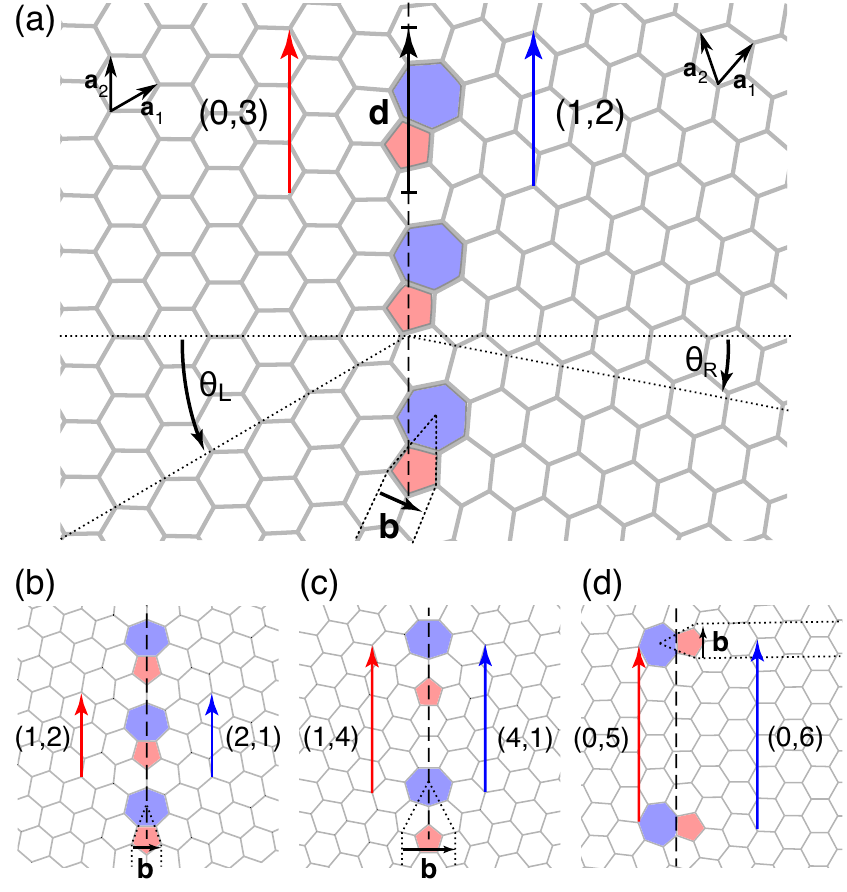}
\caption{(Color online) (a) A generic example of asymmetric 
periodic grain boundary composed of $\mathbf{b}=(1,0)$ dislocations. 
This grain-boundary structure is characterized by 
rotation angles $\theta_{\mathrm{L}}=30^{\circ}$ and $\theta_{\mathrm{R}}=10.9^{\circ}$, and a pair of matching vectors $(0,3)|(1,2)$.  
The periodicity vector $\mathbf{d}$ and the Burger's vector $\mathbf{b}$ are shown. 
Structures of symmetric grain boundaries formed by (b) $\mathbf{b}=(1,0)$ and (c) $\mathbf{b}=(1,1)$ dislocations.
(d) Degenerate grain boundary ($\theta=0^{\circ}$) with the Burger's vector of constituent 
dislocations oriented along the grain-boundary line (shown as dashed line).
 }
\label{Fig1}
\end{figure}

We consider grain-boundary models constructed as 
periodic arrays of dislocations following the Read-Shockley model
\cite{Read50}. Only dislocations formed by pentagons and heptagons 
are investigated as these structures preserve the three-fold coordination
of $sp^2$ carbon atoms thus ensuring energetically favorable configurations
of defects. This construction is consistent with experimental atomic
resolution images of grain boundaries in polycrystalline graphene
\cite{Huang11,Kim11}. The relative positions of pentagons and
heptagons defines the Burger's vectors of constituent 
dislocations. The Burger's vectors, their orientation with respect 
to the grain boundary line, and the distance between dislocation
cores define the grain boundary's structural topological invariant---the 
misorientation angle $\theta = \theta_{\rm L} + \theta_{\rm R}$. 
This relation allows constructing
arbitrary grain boundary models, as described in Ref.~\onlinecite{Yazyev10a}. Alternatively, periodic grain 
boundaries
can be defined in terms of a pair of matching vectors $(n_{\rm L}, m_{\rm L})|(n_{\rm R}, m_{\rm R})$, introduced in 
Ref.~\onlinecite{Yazyev10b} [Fig.~\ref{Fig1}(a)]. In this paper, 
however, we shall 
constrain our discussion to the Burger's vectors $\mathbf{b}$ and the
inter-dislocation distances $d$ in order to simplify the discussion.
Figure~\ref{Fig1}(a) shows a generic example of an asymmetric grain 
boundary formed by the $\mathbf{b}=(1,0)$ dislocations. 
Figures~\ref{Fig1}(b) and \ref{Fig1}(c) depict examples of symmetric 
($\theta_{\rm L}=\theta_{\rm R}$) periodic grain boundaries formed 
by $\mathbf{b}=(1,0)$ and $\mathbf{b}=(1,1)$ dislocations, 
respectively. 
Figure~\ref{Fig1}(d) shows an example of a degenerate grain 
boundary ($\theta=0^{\circ}$) with the Burger's vector of constituent 
dislocations oriented along the grain boundary line.
In our study, we focus only on the models that do not result in transport gaps due to selection by momentum \cite{endnote2}. 

\begin{figure}[b]
\includegraphics[width=86mm]{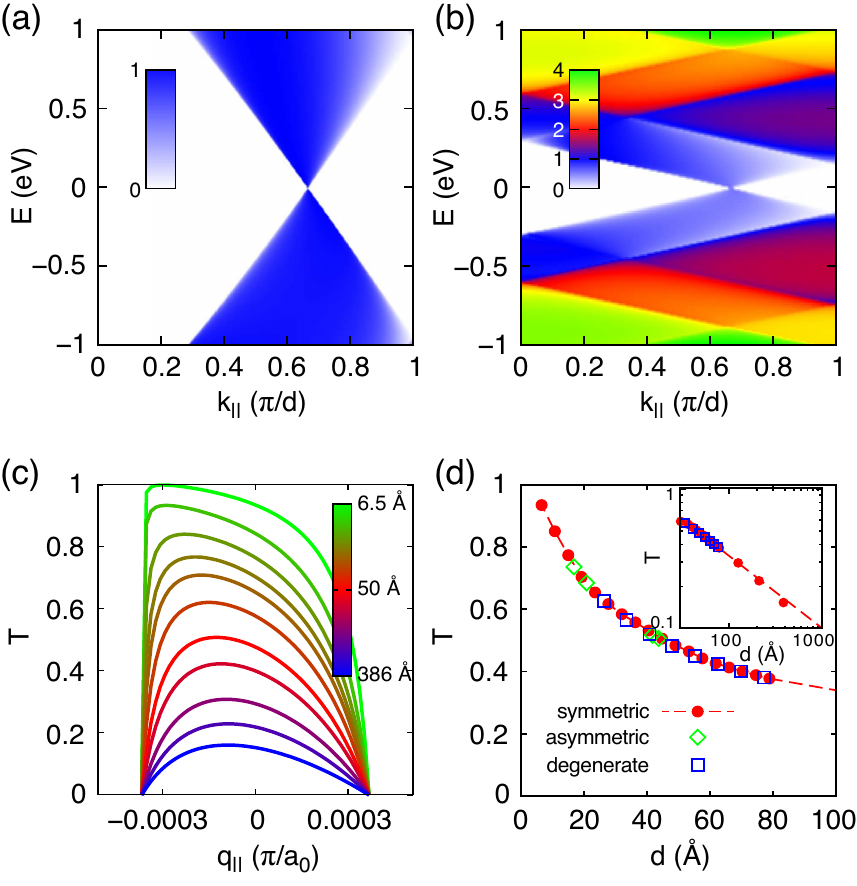}
\caption{(Color online) Electronic transport across periodic grain boundaries in graphene formed by the $\mathbf{b}=(1,0)$ dislocations.
(a),(b) Transmission probability as a function of energy $E$ and transverse momentum $k_{\parallel}$ across symmetric grain boundaries characterized by $d=6.51$~\AA\ and $d=36.2$~\AA, respectively.
(c) Transmission probability close to the Dirac point ($E=10^{-3}t$) as a function of $q_{\parallel}$ for different values of $d$. 
(d) Low-energy transmission of the normally-incident charge carriers
as a function of inter-dislocation distance $d$ for symmetric, asymmetric and degenerate grain boundaries. The inset shows the logarithmic scale plot.
}
\label{Fig2}
\end{figure}

We first focus on symmetric periodic grain boundaries formed by 
${\mathbf b}=(1,0)$ (=2.46~\AA) dislocations. Such grain boundaries are 
defined by pairs of matching vectors belonging to the $(l,l+1)|(l+1,l)$ 
series ($l \in \mathbb{N}$). Hence, $d = a_0 \sqrt{3l(l+1)+1}$ 
where $a_0=2.46$~\AA\ is the
lattice constant of graphene. Figure~\ref{Fig2}(a) shows the 
transmission probability $T$ as a function of energy $E$ and 
transverse momentum $k_{\parallel}$ for the first member of this 
sequence ($l=1$) characterized by $d=6.51$~\AA\ [Fig.~\ref{Fig1}(b)].
One clearly observes a projected Dirac cone in the irreducible half 
of the one-dimensional Brillouin zone corresponding to the periodic 
grain-boundary structure with $T(k_\parallel,E) \lesssim 1$, in 
agreement with previous calculations \cite{Yazyev10b}.
Figure~\ref{Fig2}(b) shows $T(k_\parallel,E)$ for a grain boundary
characterized by $l=8$ and, hence, a larger periodicity $d=36.2$~\AA.
The most evident difference between the two is the occurrence of multiple conductance
channels as a result of band folding over smaller Brillouin zone.
The striking feature, however, is the clear reduction of conductance
close to the Dirac point energy $E=0$. The counter-intuitive decrease of
transmission or, equivalently, enhancement of scattering upon
decreasing the density of dislocations suggests the topological
origin of the observed transport behavior. Figures~\ref{Fig2}(c) and \ref{Fig2}(d) 
further investigate the details of charge-carrier transmission 
at very low energy ($E=10^{-3}t$). One clearly observes a 
monotonic decrease of $T(q_{\parallel})$ (with 
$q_{\parallel}=k_{\parallel} - (2\pi)/(3d)$ being the transverse 
momentum relative to the location of the 
projected Dirac point) as $d$ 
increases [Fig.~\ref{Fig2}(c)]. The transmission probability of 
normally incident charge carriers ($q_{\parallel}=0$) exhibits
an inverse power scaling law $T \propto d^{-\gamma}$, with an exponent 
$\gamma \approx 0.5$ [dashed lines in Fig.~\ref{Fig2}(d)]. 
Moreover, the observed scaling law is independent of the orientation
of the Burger's vectors of dislocations relative to the 
grain-boundary line. This was explicitly demonstrated using several 
models of asymmetric and degenerate grain boundaries [Fig.~\ref{Fig2}(d)].   

\begin{figure}
\includegraphics[width=86mm]{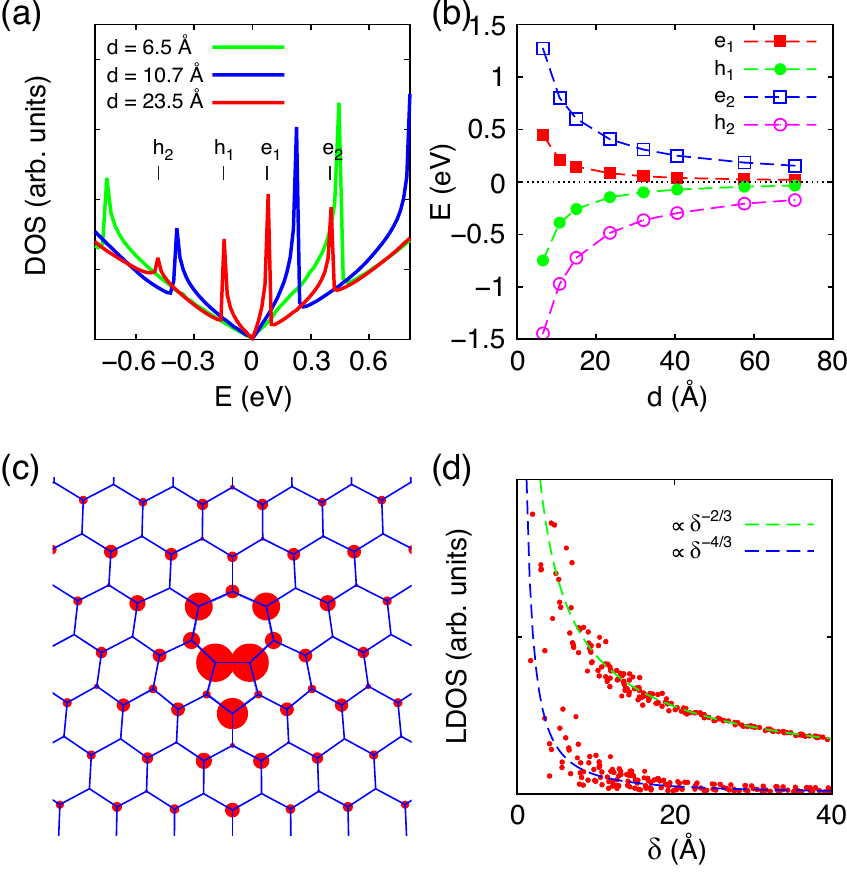}
\caption{(Color online) (a) Density of states (DOS) in 
the interface region computed for the grain boundaries formed 
by ${\mathbf b}=(1,0)$
dislocations with different values of $d$. (b) Positions of the DOS 
peaks as a function of $d$. The labels refer to peaks in panel (a) shown 
for the $d = 23.5$~\AA\ grain boundary.  
(c) Local density of states (LDOS) at 
$E=10^{-3}t$ for the $d=61.8$~\AA\ grain boundary. Circle 
areas are proportional to the LDOS.
(d) LDOS as a function of distance from the defect core $\delta$ 
calculated at $E=10^{-3}t$ for the $d= 385$~\AA\ grain boundary. 
Solid lines indicate two trends consistent with the results 
of Ref.~\onlinecite{Mesaros10}.}
\label{Fig3}
\end{figure}

The observed transport anomaly is further investigated
by analyzing the local density of states (LDOS) calculated as
\begin{equation}
\label{LDOS}
 \mathrm{LDOS}_{n}\left(E\right)=-\frac{d}{\pi^2} \int_0^{\pi/d} Im\left(G_{\mathrm{S}}\left(E,k_\parallel \right)\right)_{n,n} dk_\parallel .
\end{equation}
Figure~\ref{Fig3}(a) shows the density of states (DOS) calculated by 
summing the LDOS over atoms located in the grain-boundary region of
40~\AA\ width. One clearly observes the presence of sharp van Hove 
singularities both in the valence and conduction bands 
superimposed on the linear contribution 
of pristine graphene. The peak positions converge to the Dirac point 
energy as the distance between dislocations $d$ increases 
[Fig.~\ref{Fig3}(b)]. These DOS peaks can be attributed to the
electronic states localized at the dislocations, as corroborated by 
Figure~\ref{Fig3}(c). The presence of localized states results in 
resonant back-scattering at low energies, similar to 
dopants and covalent functionalization defects in graphene \cite{Choi00,Titov10,Wehling10,Yuan10,Ferreira11b,Radchenko12}.

The origin of the localized states is related to the
topological nature of defects in polycrystalline graphene. Charge 
carriers of momentum ${\mathbf k}$ encircling a dislocation with
Burger's vector ${\mathbf b}$ gain a phase $\varphi = {\mathbf k} \cdot {\mathbf b}$ \cite{Iordanskii86}. The aforementioned 
${\mathbf b} = (1,0)$ dislocations thus give rise to 
$\varphi = + 2\pi/3$ and $\varphi = - 2\pi/3$ for the charge 
carriers in valleys $\tau = +1$ and $\tau = -1$, respectively.
Starting from the Dirac equation for a massless particle
\begin{equation}
\label{Gauge}
H = \left(p_x - i A_x \right) \sigma_x + \left(p_y - i A_y \right) \tau \sigma_y ,
\end{equation}
the effect
of a dislocation is accounted for by means of a gauge field ${\mathbf A}\propto\mathbf{k}\cdot\mathbf{b}$ \citep{Lammert00,Cortijo07,Mesaros09,Vozmediano10}.
Using this continuum model, Mesaros {\it et al.} predicted that
an isolated ${\mathbf b} = (1,0)$ dislocation gives rise to quasi-localized
modes at $E=0$ \cite{Mesaros10}. 
The continuous model for $\mathbf{b}=(1,0)$ has two low-energy 
solutions with the LDOS decaying as $\propto\delta^{-2/3}$ and
$\propto\delta^{-4/3}$, where $\delta$ is the 
distance from the defect core. Our numerical calculations for the 
grain-boundary model with a very large distance between dislocations 
confirm
the analytical result showing the two solutions coexisting
on different sublattices of the graphene lattice [Fig.~\ref{Fig3}(d)].

In order to gain a qualitative understanding of the dependence of
transmission on $d$ [Figs.~\ref{Fig2}(c) and \ref{Fig2}(d)], one has to appreciate the
fact that the finite distance between dislocations forming a grain
boundary allows for the hybridization of 
localized states. As a result, the LDOS peaks at positive and 
negative energies emerge in lieu of the $E=0$
peak for an isolated dislocation. As the distance between 
dislocations $d$ increases, the hybridization diminishes, thus 
reducing the peak energies [Figs.~\ref{Fig3}(a) and \ref{Fig3}(b)] and resulting in
the progressive decrease of the transmission 
close to $E=0$. At finite energies, however, the minimum of 
transmission is achieved at a certain distance between 
dislocations. It is worth stressing that despite the fact that our
conclusions are based on periodic  models of dislocations, 
there is no strict requirement of periodicity, in contrast to the 
case of suppressed conductivity due to momentum conservation 
\cite{Yazyev10b}. This has been explicitly verified by means of 
supercell
calculations which show that transmission is 
insensitive to perturbation of the periodic arrangement of 
dislocations in grain boundaries.

\begin{figure}
\includegraphics[width=86mm]{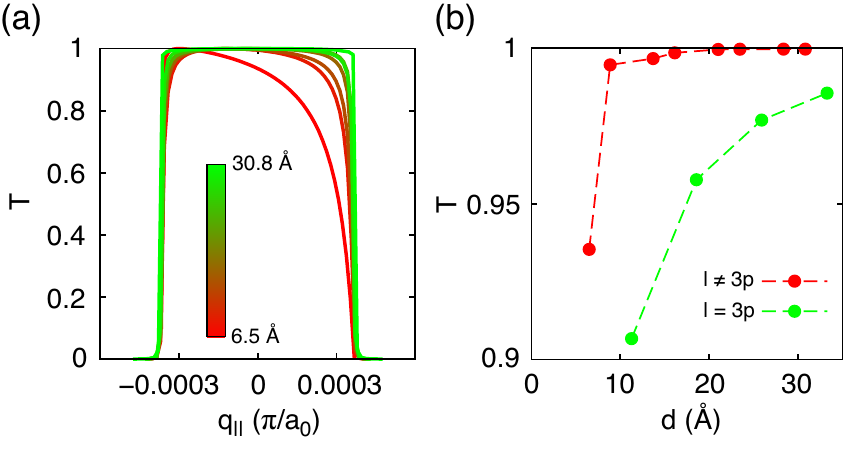}
\caption{(Color online) 
Electronic transport across grain boundaries  composed of 
$\mathbf{b}=(1,1)$ dislocations.
(a) Transmission probability close to the Dirac point ($E=10^{-3}t$)
as a function of $q_{\parallel}$ for grain boundaries 
characterized by different values of $d$ within the $l \neq 3p$ family.
(b) Low-energy transmission of the normally-incident charge carriers
as a function of inter-dislocation distance $d$. Two families of 
grain boundaries are distinguished.
}
\label{Fig4}
\end{figure}

More generally, all dislocations characterized by Burger's vectors
${\mathbf b}=(n,m)$ with $n-m \neq 3q$ ($q \in \mathbb{Z}$) have a similar
effect on charge carriers in graphene because of equal values for the
${\mathbf k} \cdot {\mathbf b}$ product for the two valleys. 
However, dislocations with $n-m=3q$
are expected to behave as ordinary (topologically trivial) 
scatterers since ${\mathbf k} \cdot {\mathbf b} = 0$.
We verify this statement by investigating the transmission through the 
grain boundaries formed by ${\mathbf b}=(1,1)$ ($=4.23$~\AA) dislocations 
[Fig.~\ref{Fig1}(c)]. Following the convention defined earlier, 
these grain boundaries are defined by
the pairs of matching vectors belonging to $(1,l+1)|(l+1,1)$ series 
($l \in \mathbb{N}$). Figure~\ref{Fig4}(a) shows that already the
first members of this family exhibit transmission probabilities 
close to 1, which further increase as the inter-dislocation distance  
$d$ increases. Moreover, one can distinguish two families of 
grain-boundary structures characterized by $l = 3p$ and $l \neq 3p$ ($p \in \mathbb{N}$) [Fig.~\ref{Fig4}(b)]. Within both families the 
effect of dislocations can be described in terms of scattering cross-sections, $\sigma_{l = 3p} \approx 0.5$~\AA\ 
and $\sigma_{l \neq 3p} \approx 0.01$~\AA. The first value is 
significantly larger since both Dirac points correspond to 
$k_\parallel =0$, thus enabling the intervalley scattering process 
upon transmission. The LDOS calculated for these grain boundary 
configurations show no localized states at low energies
(not shown here). 

To conclude, our study reveals an intriguing aspect 
of charge-carrier transport in topologically disordered graphene. 
Predicted anomalous scattering is 
especially pertinent to low-angle grain boundaries 
($d \gg |{\mathbf b}|$) composed of dislocations with the minimal 
Burger's vector ${\mathbf b}=(1,0)$. These dislocations
are dominant in realistic samples due to the reduced
elastic response \cite{Nelson02}, and may even occur within 
seemingly single-crystalline domains of graphene \cite{Coraux08}.
Unlike covalently bound adatoms which also act as 
resonant scattering centers \cite{Titov10,Wehling10,Yuan10,Ferreira11b,Radchenko12}, 
dislocations cannot be easily eliminated from the sample
due to their topological nature and high diffusion barriers
at normal conditions \cite{Banhart10}.
Our work thus identifies an important source of charge-carrier scattering
in large-area samples of graphene produced by high 
efficiency techniques, such as the chemical vapor deposition. 

We would like to thank G.~Aut\`es, A.~Cortijo, M.~I.~Katsnelson,
L.~S.~Levitov, and A.~Mesaros for discussions. This work was
supported by the Swiss National Science Foundation (Grant No. PP00P2\_133552). Computations have been performed at the Swiss National Supercomputing Centre (CSCS) under project s443.


\begin{references}
\bibitem{endnote1}
We verified by means of explicit calculations that including the 
dependence of hopping integrals on the variations of interatomic
distance due to the elastic strain fields produced by the 
dislocations has only minor effect on the low-energy charge-carrier
transmission. The tight-binding results also agree with the results
of first-principles calculations as shown previously \cite{Yazyev10b}.
\bibitem{endnote2}
Only periodic structures characterized by marching vectors 
$(n_{\rm L}, m_{\rm L})|(n_{\rm R}, m_{\rm R})$ such that either 
both $n_{\rm L} - m_{\rm L} = 3p$ and $n_{\rm R} - m_{\rm R} = 3q$,
or both $n_{\rm L} - m_{\rm L} \ne 3p$ and $n_{\rm R} - m_{\rm R} \ne 3q$
($p,q \in \mathbb{Z}$) are considered \cite{Yazyev10b}.
\end{references}
\end{document}